\documentclass[pra,twocolumn,amsmath,amssymb,eqsecnum,superscriptaddress,longbibliography]{revtex4-1}

\usepackage{graphicx,upgreek,color,mathtools,bm,mathptmx,wasysym,mathptmx}
\usepackage[hyphenbreaks]{breakurl}
\usepackage[colorlinks=true,linkcolor=blue,citecolor=blue,urlcolor =blue]{hyperref}

\newcommand{\Tr}{\mathop{\mathrm{Tr}} \nolimits}

\newcommand{\I}{\mathop{\mathrm{Im}}\nolimits}

\newcommand{\ML}{\{ \! \! \! \{}
\newcommand{\MR}{\} \! \! \! \}}
\begin{document}

\title{Quasiprobability currents on the sphere}

\author{Iv\'an~F.~Valtierra}
\affiliation{Departamento de Fisica, Universidad de
  Guadalajara, 44420~Guadalajara, Jalisco, Mexico}

\author{Andrei~B.~Klimov}
\affiliation{Departamento de Fisica, Universidad de
  Guadalajara, 44420~Guadalajara, Jalisco, Mexico}

\author{Gerd~Leuchs}
\affiliation{Max-Planck-Institut f\"ur die Physik des Lichts,
  Staudtstra{\ss}e 2, 91058~Erlangen, Germany}

\author{Luis~L.~S\'anchez-Soto}
\affiliation{Max-Planck-Institut f\"ur die Physik des Lichts,
  Staudtstra{\ss}e 2, 91058~Erlangen, Germany}
\affiliation{Departamento de \'Optica, Facultad de Fisica,
  Universidad Complutense, 28040~Madrid, Spain}

\begin{abstract}
  We present analytic expressions for the $s$-parametrized currents on the sphere for both unitary and dissipative evolutions. We examine the spatial distribution of the flow generated by these currents for quadratic Hamiltonians. The results are applied for the study of the quantum dissipative dynamics of the time-honored Kerr and Lipkin models, exploring the appearance of the semiclassical limit in stable, unstable and tunnelling regimes.
\end{abstract}

\maketitle

\section{Introduction}

Presenting quantum mechanics as a statistical theory on a classical phase space has attracted a great deal of attention since the very early days of this discipline. This framework gives an alternative point of view that provides more insight and understanding~\cite{Hillery:1984aa,Lee:1995aa,Schroek:1996aa,Ozorio:1998aa,Schleich:2001aa,QMPS:2005aa}, and, in addition, avoids the operator formalism, thereby freeing quantization of the burden of the Hilbert space~\cite{Ali:2005aa}.

The main ingredient for any successful phase-space method is a \emph{bona fide} mapping that relates operators with functions defined on a smooth manifold $\mathcal{M}$, the phase space of the system, endowed with a very precise mathematical structure~\cite{Kirillov:2004aa}. This mapping, first suggested by Weyl~\cite{Weyl:1928aa} and later put on solid grounds by Stratonovitch~\cite{Stratonovich:1956aa}, is not unique. In fact, a whole family of $s$-parametrized functions can be assigned to each operator and the choice of a particular element of the family depends on  its convenience for each problem. In particular, the time-honored quasiprobability distributions are the functions connected with the density operator. The most common choices of $s$ are $+1$, 0, and $-1$, which correspond to the $P$ (Glauber-Sudarshan)~\cite{Glauber:1963aa,Sudarshan:1963aa}, $W$~(Wigner)~\cite{Wigner:1932uq}, and $Q$~(Husimi)~\cite{Husimi:1940aa} functions, respectively. For continuous variables (such as Cartesian position and momentum), the quintessential example that fuelled the interest for this field, the parameter $s$ defines different orderings of the basic variables.

These quasiprobability distributions and their corresponding equations of motion give the right tools for the representation of quantum dynamics entirely in the language of phase-space variables. Actually, a substantial step to addressing this question came from the work of Groenewold~\cite{Groenewold:1946aa} and Moyal~\cite{Moyal:1949aa}, who showed that the evolution equation can be written as 
\begin{equation}
  \partial_{t}W_{\varrho}^{(s)}(\Omega, t) =
  \ML W_{\varrho}^{(s)}(\Omega, t), W_{H}^{(s)}(\Omega ) \MR  \, ,
  \label{ee1}
\end{equation}
where $\Omega \in \mathcal{M}$ are points in phase space, $W_{H}^{(s)}(\Omega )$ is the Weyl symbol of the Hamiltonian and  the Moyal bracket $\ML \cdot, \cdot \MR$ is the image of the  commutator [times $(i \hbar)^{-1}$] under the Weyl-Stratonovitch map. Therefore, (\ref{ee1}) is formally identical with its quantum version for the density operator, if one replaces the commutator with the Moyal bracket. This equation contains,  in general, higher-order derivatives, which reflects the analytical properties of the map. This complicates, in general, getting any analytical solution~\cite{Polkovnikov:2010aa}.

It turns out that the dynamics in (\ref{ee1}) can be rewritten as a continuity equation
\begin{equation}
  \partial_{t}W_{\varrho}^{(s)}(\Omega, t)= -
  \nabla \cdot \mathbf{J}^{(s)}(\Omega , t) \, ,
  \label{flow}
\end{equation}
as it was recently discussed for one-dimensional systems~\cite{Bauke:2011aa,Steuernagel:2013aa,Kakofengitis:2017aa,Kakofengitis:2017ab,Oliva:2019aa,Oliva:2019ab}. Nonetheless, there is a substantial difference between the classical and quantum regimes. 

In classical statistical mechanics, the conservative evolution of the (positive) phase-space distribution function $f(\Omega, t)$ is given by the famous Liouville equation~\cite{Pathria:2011aa}
\begin{equation}
\label{eq:liouv}
\partial_{t} f(\Omega,t ) = \{ f(\Omega, t), H_{\mathrm{cl}} \} \, ,
\end{equation}
with $\{ \cdot , \cdot \}$ being the Poisson brackets on $\mathcal{M}$. This can also be expressed as a continuity equation
\begin{equation}
\label{eq:contclass}
\partial_{t}f (\Omega, t)= - 
\nabla \cdot \mathbf{J}_{\mathrm{cl}}(\Omega , t) \, ,
\end{equation}
but now the classical current is
\begin{equation}
\label{eq:Jcl}
\mathbf{J}_{\mathrm{cl}} = \mathbf{v} (\Omega) \, f(\Omega, t) \
\end{equation}
where $\mathbf{v}(\Omega )\propto \nabla H_{\mathrm{cl}}$ is related to the velocity field generated by the Hamiltonian $H_{\mathrm{cl}}$ on $\mathcal{M}$. In other words, $\mathbf{v}(\Omega )$ induces a Hamiltonian propagation 
\begin{equation}
f(\Omega ,0)\overset{\mathbf{v}}{\longmapsto } 
f(\Omega ,t)=f(\Omega_{\mathrm{cl}}(t),0) \, ,  
\label{ft}
\end{equation}
so that the distribution function is preserved along every classical trajectory $\Omega _{\mathrm{cl}} (t)$. The  stagnation points $\{\Omega _{j}\}$, where $\mathbf{J}_{\mathrm{cl}}(\Omega_{j},t)=0$, coincide either with the zeros of $\mathbf{v}(\Omega)$\ or those of $f(\Omega , 0)$.

For the quantum dynamics, $\mathbf{J}^{(s)}(\Omega ,t)$ can be conveniently represented as a (differential) operator acting on the quasiprobability distribution; i.e.,
\begin{equation}
\mathbf{J}^{(s)}(\Omega ,t) = \hat{\mathbf{J}}^{(s)}(\Omega )\;
W_{\varrho}^{(s)}(\Omega ,t)\,. 
\label{Jsin}
\end{equation}
In the semiclassical limit, when the dynamics of distribution resembles the motion of an incompressible fluid, $\hat{\mathbf{J}}^{(s)}(\Omega ) \rightarrow  \mathbf{v} (\Omega )$. In contradistinction, in the full quantum regime trajectories do not exist globally for systems whose phase-space distributions can develop areas with nonpositive values and the velocity is not a well-defined function~\cite{Oliva:2018aa}. Consequently, (\ref{Jsin}) not only describes the propagation and deformations of $W_{\varrho }^{(s)}(\Omega ,t)$, as in (\ref{ft}), but also the emergence of interference patterns (in particular, the evolution of $W_{\varrho }^{(s)}(\Omega )$ into a zero-amplitude function at the nonclassical stagnation points).

The currents $J^{(s)}(\Omega ,t)$ bear several unexpected features from a classical viewpoint~\cite{Steuernagel:2013aa,Kakofengitis:2017aa,Kakofengitis:2017ab,Oliva:2019aa,Oliva:2019ab}.  The distribution and character of the stagnation points can be used for detecting the quantumness of the evolution, both from the Wigner and the Husimi currents~\cite{Veronez:2016aa}.

Interestingly, the quasiprobability currents can be properly defined not only in the unitary case, but also in the presence of dissipation~\cite{Kossakowski:1972aa,Lindblad:1976aa,Gorini:1976aa}. The analysis of these dissipative currents can provide interesting insights into the decoherence dynamics~\cite{Braasch:2019aa}, especially in the presence of purely quantum effects, such as self-interference and tunnelling.

Recently, it has been shown that the description of the evolution in terms of currents can be extended in natural ways to spinlike systems, where the classical phase space is the unit sphere~\cite{Yang:2019aa}. Actually, in the simplest yet relevant case of a nonlinear Kerr medium, the spatial distribution of the phase-space Wigner current and in, particular, of the stagnation lines (one of the components of the Kerr current is always zero) allows one to distinguish quantum from classical dynamics, even for short times.

In the present paper we introduce the $s$-ordered currents for spinlike systems whose evolution is governed by quadratic Hamiltonians and apply it to the analysis of the dynamics of the Kerr~\cite{Kitagawa:1993aa, Agarwal:1997aa} and Lipkin-Meshkov-Glick~\cite{Lipkin:1965aa,Meshkov:1965aa,Glick:1965aa} models, both in the unitary and the dissipative case. The most representative examples of spin dynamics, stable, unstable and tunnelling (evolution in the classically forbidden regions), will be analyzed on the basis of spatial distributions of the quantum currents.   Although we shall get analytical expressions for any $s$-ordered distribution, our numerical analysis will be focused on the Wigner propagation, 
for it reveals in the most conspicuous manner the quantum dynamical behavior in case of nonlinear evolution.

\section{Quasiprobability distributions on the sphere}

We consider a system whose dynamical symmetry group is SU(2).  The corresponding Lie algebra $\mathfrak{su} (2)$ is spanned by the operators $\{ \hat{S}_{x}, \hat{S}_{y}, \hat{S}_{z} \}$ satisfying the standard commutation relations $ [\hat{S}_{x}, \hat{S}_{y} ] = i  \hat{S}_{z}$ and cyclic permutations (in units $\hbar =1$, which will be used throughout). The Casimir operator is
$\hat{\mathbf{S}}^{2} = \hat{S}_{x}^{2} + \hat{S}_{y}^{2}+ \hat{S}_{z}^{2} = S (S+1) \openone$, so the eigenvalue $S$ (which is a nonnegative integer or half integer) labels the irreducible representations (irreps).

We take a fixed irrep of spin $S$, with a $(2S+1)$-dimensional carrier space $\mathcal{H}_{S}$ spanned by the standard angular momentum basis $\{ |S, m\rangle \mid m= -S, \ldots, S\}$, whose elements are simultaneous eigenstates of $\hat{\mathbf{S}}^{2}$ and $ \hat{S}_{z}$:
\begin{equation}
  \hat{\mathbf{S}}^{2} |S, m \rangle = S (S+1) |S, m \rangle \, ,
  \qquad
  \hat{S}_{z} |S, m\rangle = m |S, m \rangle \, .
\end{equation}

The highest weight state is $|S,S\rangle$ and it is annihilated by the ladder operator $\hat{S}_{+}$ (with $\hat{S}_{\pm} = \hat{S}_{x} \pm i  \hat{S}_{y}$). The isotropy subgroup (i.e., the largest subgroup that leaves the highest weight state invariant) consists of all the elements of the form $\exp( i \chi \hat{S}_{z})$, so it is isomorphic to U(1). The coset space is then SU(2)$/$U(1), which is just the unit sphere $\mathcal{S}_{2}$ (the so-called Bloch sphere) and it is the classical phase space, the natural arena to describe the dynamics.

The $s$-parametrized Weyl-Stratanovich map
\begin{equation}
  \hat{A}  \mapsto  W_{A}^{(s)} (\Omega ) =
  \Tr [  \hat{A} \, \hat{w}^{(s)}(\Omega ) ]  ,
  \label{map}
\end{equation}
where $\Omega =(\theta ,\phi )\in \mathcal{S}_{2}$,  puts in one-to-one correspondence each operator $\hat{A}$ invariantly acting on $\mathcal{H}_{S}$ with a function on the sphere $\mathcal{S}_{2}$. The corresponding kernels  $\hat{w}^{(s)}$ are defined as~\cite{Berezin:1975mw,Agarwal:1981bd,Varilly:1989ud}   
\begin{equation}
  \hat{w}^{(s)}(\Omega )= \sqrt{\frac{4\pi}{2S+1}}
  \sum_{K=0}^{2S} \sum_{q=-K}^{K}
  ( C_{SS,K0}^{SS} )^{-s} \, Y_{Kq}^{\ast}(\Omega)
  \hat{T}_{Kq}^{S}\,,
  \label{ker}
\end{equation}
where $Y_{Kq}(\Omega )$ are the spherical harmonics, $C_{S_{1}m_{1},S_{2}m_{2}}^{Sm}$ the Clebsch-Gordan coefficients and $\hat{T}_{Kq}^{S}$ the irreducible tensor operators~\cite{Fano:1959ly,Varshalovich:1988ct} 
\begin{equation}
  \hat{T}_{Kq}^{S} = \sqrt{\frac{2K+1}{2S+1}}
  \sum_{m,m^{\prime}=-S}^{S} C_{Sm,Kq}^{Sm^{\prime}} \,
  |S,m^{\prime}\rangle \langle S,m| \, .
\end{equation}
As expected, they are properly normalized 
\begin{equation}
  \Tr [ \hat{w}^{(s)} ( \Omega ) ]=1 \, ,
  \qquad
  \frac{2S+1} {4\pi}\int_{\mathcal{S}_{2}}
  d\Omega \; \hat{w}^{(s)} (\Omega )=\openone \,,
\end{equation}
with $d\Omega =\sin \theta\ d\theta\ d\phi $ the invariant measure on the sphere.

Consequently, the symbol of $\hat{A}$ can be concisely  expressed as
\begin{equation}
  W_{A}^{(s)} ( \Omega ) = \sqrt{\frac{4\pi}{2S+1}}
  \sum_{K=0}^{2S} \sum_{q=-K}^{K}  ( C_{SS,K0}^{SS} )^{-s}
  A_{Kq} \; Y_{LM}^{\ast} (\Omega ) \, ,
\label{eq:symb}
\end{equation}
where $A_{Kq}=\Tr ( \hat{A} \hat{T}_{Kq}^{S \dagger} )$.  As some relevant examples we shall need in what follows we quote  
\begin{equation}
\begin{array}{rcl}
  \label{eq:symbols}
  \hat{S}_{i} & \mapsto & W^{(s)}_{S_{i}}
=  \displaystyle \left ( \frac{S}{S+1} \right )^{-s/2}
\sqrt{S(S+1)} \,   n_{i} \, ,  \\
  & & \\
  \{\hat{S}_{i},\hat{S}_{j}\}_{+} & \mapsto &
W^{(s)}_{\{S_{i},S_{j}\}_{+}} =  \mathcal{C}_{ij}^{(s)}  \; n_{i}n_{j}  +   \frac{1}{3}\delta_{ij} [ 2S(S+1) - 
\mathcal{C}_{ij}^{(s)} ] , 
\end{array}
\end{equation}
where $\mathbf{n} $ is a unit vector in the direction of $\Omega \in \mathcal{S}_{2}$, $\{ \cdot, \cdot \}_{+}$ stands for the anticommutator and $\mathcal{C}_{ij}^{(s)}=(1-\frac{1}{2}\delta_{ij})[S(2S-1)]^{(1-s)/2}[(2S+3)(S+1)]^{(1+s)/2}$.

The traditional SU(2) quasiprobability distributions are just the $s$-symbols of the density operator $\hat{\varrho}$. The value $s=0$ corresponds to the standard Wigner function, whereas $s=\pm 1$ leads to $P$ and $Q$ functions respectively, defined as dual coefficients in the basis of spin coherent states~\cite{Arecchi:1972aa,Perelomov:1986ly}
\begin{equation}
|\Omega \rangle = \exp [ \case{1}{2} \theta
  (\hat{S}_{+}e^{-i\phi}- \hat{S}_{-}e^{i\phi} ) ]
  |S,S\rangle \, ,
\label{CS}
\end{equation}
according to
\begin{equation}
  Q(\Omega ) = \langle \Omega |\hat{\varrho}|\Omega \rangle \, ,
  \qquad
    \hat{\varrho} =\frac{2S+1}{4\pi}\int_{\mathcal{S}_{2}}d\Omega \;
  P(\Omega) \; |\Omega \rangle \langle \Omega | \, .
\end{equation}
The symbols $W_{A}^{(s)} (\Omega )$ are covariant under SU(2) transformations and provide the overlap relation
\begin{equation}
  \Tr (\hat{\varrho} \hat{A}) = \frac{2S+1}{4\pi}
  \int_{\mathcal{S}_{2}} d\Omega \, W_{\varrho}^{(s)}(\Omega )
  \,W_{A}^{(-s)}(\Omega ) \, .
\end{equation}
A number of alternative generalized quasiprobability distributions can be found using the method of Cohen~\cite{Cohen:1966aa} (see also Ref.\cite{Liu:2011ab}).

In this representation, the Moyal equation (\ref{ee1}), as indicated in the Introduction, involves higher-order derivatives. However, it admits an expansion on the parameter $\varepsilon =(2S+1)^{-1}$. When $\varepsilon \ll 1$ we are in the semiclassical limit and one can show that~\cite{Gilmore:1975aa,Amiet:1991aa,Benedict:1999aa,Klimov:2005aa}
\begin{equation}
  \partial_{t}W_{\varrho}^{(s)}(\Omega, t ) \simeq 2 \varepsilon
  \{W_{\varrho}^{(s)} (\Omega, t ), W_{H}^{(s)}(\Omega )\} + s \mathcal{O}(\varepsilon^{2})+
\mathcal{O}(\varepsilon^{3}),
\end{equation}
where $\{\cdot, \cdot \}$ are the Poisson brackets on the sphere $\mathcal{S}_{2}$
\begin{equation}
  \{f (\Omega ),g(\Omega )\} = \frac{1}{\sin \theta}
  ( \partial_{\phi}f \; \partial_{\theta}g -
  \partial_{\theta}f \; \partial_{\phi}g ) .
\label{PB}
\end{equation}
Therefore, the first-order corrections to the classical evolution $\sim \mathcal{O}(\varepsilon^{2})$ vanish for the Wigner function and consist of second-order derivatives of $W_{\varrho}^{(\pm 1)}(\Omega )$.

The lowest order approximation, known as the Truncated Wigner Approximation (TWA)~\cite{Heller:1976aa,Davis:1984aa,Bagrov:1992aa,Drummond:1993aa,Polkovnikov:2003aa,Drummond:2017aa,Valtierra:2017aa}, describes propagation of every point of the initial distribution along the corresponding classical trajectories $\Omega (t)$, which are solutions of the Hamilton equations; viz,
\begin{equation}
  W_{\varrho}^{(s)}(\Omega, t) \simeq
  W_{\varrho}^{(s)}(\Omega(-t), t=0) \, .
  \label{TWA}
\end{equation}
This semiclassical evolution allows one to predict the short-time behavior~\cite{Valtierra:2016aa,Sundar:2019aa}. The positive and negative parts of $W_{\varrho }^{(s)}(\Omega ,t=0)$  are deformed according to (\ref{TWA}), so that their volumes are preserved during the validity of the TWA.

\section{Hamiltonian dynamics and currents on the sphere}

The exact evolution equation for $W_{\varrho}^{(s)}(\Omega )$ has been derived in \cite{Klimov:2002cr} (see also \cite{Zueco:2007aa} and \cite{Klimov:2017aa}, where the corresponding star-product for the map is discussed). In most physical applications only Hamiltonians quadratic in the spin generators play an important role. Typical examples include second-harmonic generation, homogeneous spin-spin interactions, spin-orbit splitting, and atom-field interactions in the dipole approximation~\cite{Cohen-Tannoudji:1989aa}. The generic second-order Hamiltonian reads
\begin{equation}
\hat{H} = \sum_{i}a_{i} \, \hat{S}_{i}+\sum_{jk}b_{jk} \, \{ \hat{S}_{j}, 
\hat{S}_{k}\}_{+} \equiv \hat{H}_{L}+ \hat{H}_{NL} \   \label{H}
\end{equation}
where $\hat{H}_{L}$ and $\hat{H}_{NL}$ refer to the linear and nonlinear
parts of the Hamiltonian.

The evolution equation for the Hamiltonian (\ref{H}) can be rewritten in terms of the Poisson brackets (\ref{PB}) as follows: 
\begin{eqnarray}
  \partial _{t}W_{\varrho }^{(s)} & = &
\left( \frac{S}{S+1}\right)^{-s/2}
 \sum_{i}a_{i}\{W_{\varrho }^{(s)},n_{i}\} \nonumber \\
  & + &  \frac{1}{2\varepsilon }
\sum_{j,k}b_{jk}[\{\hat{G}_{k}W_{\varrho }^{(s)},n_{j}\}+\{\hat{G}
_{j}W_{\varrho }^{(s)},n_{k}\}]\,,
\label{EE2}
\end{eqnarray}
where 
\begin{eqnarray}
  \hat{G}_{k}^{(\pm 1)} &=&n_{k}(1\pm \varepsilon )\,\pm
i \varepsilon (\mathbf{n}\times \hat{\mathbf{L}})_{k}\,,  \nonumber \\
&& \\
\hat{G}_{k}^{(0)} &=&\frac{1}{2}n_{k}\Phi (\mathcal{L}^{2})-\frac{1}{2}
 \varepsilon ^{2}[n_{k}+
2 i  (\mathbf{n}\times
 \hat{\mathbf{L}})_{k}]\Phi^{-1}(\mathcal{L}^{2})\,. \nonumber
\end{eqnarray}
Here, $\hat{\mathbf{L}}=(\hat{L}_{x},\hat{L}_{y},\hat{L}_{z})$ are a differential realization of the angular momentum operators; viz, 
\begin{eqnarray}
  \hat{L}_{x}& = & i (\sin \phi \partial _{\theta }+
  \cot \theta \cos \phi \partial _{\phi }) \,, \nonumber \\
  \hat{L}_{y}& = & i (-\cos \phi \partial _{\theta}+
  \cot \theta \sin \phi \partial _{\phi })\,,\\
  \hat{L}_{z} & = & -i \partial_{\phi }\,, \nonumber
\end{eqnarray}
and
$\hat{\mathcal{L}}^{2}=\hat{L}_{x}^{2}+\hat{L}_{y}^{2}+\hat{L}_{z}^{2}$ is the realization of the Casimir operator on the sphere, namely
\begin{equation}
  \hat{\mathcal{L}}^{2} =-\left(
    \partial _{\theta \theta }+\cot \theta \,\partial _{\theta }+
    \frac{1}{\sin ^{2}\theta }\partial _{\phi \phi}\right) \;, 
  \label{L2}
\end{equation}
so that $\hat{\mathcal{L}}^{2}Y_{Lm}(\Omega )=L(L+1)Y_{Lm}(\Omega )$ (note that, except for a sign, it is the Laplacian operator on the sphere).  Finally, the function $\Phi $ reads
\begin{equation}
  \Phi (x)=\left[ 2-\varepsilon ^{2}(2x^{2}+1) +
    2 \sqrt{1-\varepsilon^{2}(2x^{2}+1)+\varepsilon ^{4}x^{4}}\right] ^{1/2}\,.
\end{equation}

The equation of motion (\ref{EE2}) can be immediately recast in the form (\ref{flow}). The currents (\ref{Jsin}), $\mathbf{J}^{(s)} = (J_{\theta}^{(s)},J_{\phi}^{(s)})$, can be conveniently represented as the actions of differential operators on the corresponding quasiprobability distributions:
\begin{equation}
\mathbf{J}^{(s)}(\Omega , t) = \hat{\mathbf{J}}^{(s)} \;
W_{\varrho}^{(s)}(\Omega , t) \, ,
\label{Js}
\end{equation}
with 
\begin{eqnarray}
\hat{J}_{\theta}^{(s)} & = &  \left ( \frac{1-\varepsilon}{1+\varepsilon}
\right)^{-s/2} \frac{1}{\sin \theta} \sum_{i}a_{i}\partial_{\phi}n_{i}
\nonumber \\
  & + &  
\frac{1}{2\varepsilon \sin \theta} \sum_{jk}b_{jk} [\partial_{\phi}n_{j} 
\hat{G}_{k}^{(s)} + \partial_{\phi}n_{k}\hat{G}_{j}^{(s)}] \, ,  \nonumber \\
  & & \label{Jops} \\
\hat{J}_{\phi}^{(s)} & = & - \left( \frac{1-\varepsilon}{1+\varepsilon}
\right)^{-s/2} \sum_{i}a_{i}\,\partial_{\theta}n_{i} \nonumber \\
  & - &  \frac{1}{2\varepsilon}
\sum_{jk}b_{jk} [\partial_{\theta}n_{j}\hat{G}_{k}^{(s)} +
  \partial_{\theta}n_{k}\hat{G}_{j}^{(s)}] \, .  \nonumber
\end{eqnarray}
It is worth noticing that $\mathbf{\hat{J}}^{(\pm 1)} $ are first-order operators (observe that $\mathbf{n}\times \hat{\mathbf{L}} $ is just the angular part of the gradient operator in spherical coordinates~\cite{Varshalovich:1988ct}), whereas $\mathbf{\hat{J}}^{(0)}$ contains higher-order derivatives because of the to dependence on the Casimir operator~$\hat{\mathcal{L}}^{2}$.

The current associated to the linear Hamiltonian $\hat{H}_{L}$ generates a rigid rotation of the initial distribution; i.e., the
quantum and classical currents coincide in this case.

In the general case, the quantum dynamics is described by current operators that do not reduce to a multiplication by some phase-space function, as in the classical case. This leads to a nontrivial evolution of the stagnation points $\Omega _{j}$, wherein $J^{(s)}(\Omega _{j},t)=0$, as we shall see in Sec.~\ref{sec:ex}. The properties of the vector field $J^{(s)}(\Omega ,t)$ in the vicinity of the stagnation points can be studied, e.g., with the winding number~\cite{Steuernagel:2013aa,Veronez:2016aa} 
$ I(\Omega _{j}) = \frac{1}{2\pi } \oint_{L} d\varphi$,  
where $\varphi$ is the angle between the flow and some fixed reference axis in the loop $L$. This number takes the values $I(\Omega _{j})=\pm 1$ for vortices and saddle points, correspondingly.  Nontrivial stagnation points [that do not coincide with zeros of the initial distribution and the gradient field $\nabla W_{H}^{(s)}(\Omega )$] dynamically emerge/disappear only by pairs (topological dipoles~\cite{Veronez:2016aa}) according to Poincar\'{e}-Hopf theorem.

It is worth noting that the evolution  (\ref{EE2}) can be rewritten in terms solely of the Poisson brackets with the Weyl symbol of the Hamiltonian (\ref{H})  in two instances:  for the linear case $\hat{H}_{L}$ and when the Hamiltonian is the square of an element of the su(2) algebra; i.e., up to an SU(2) rotation, it has the form
\begin{equation}
\hat{H}_{NL} = b_{z} \, \hat{S}_{z}^{2},  
\qquad \qquad 
\partial _{t} W_{\varrho }^{(s)} =  
\{\hat{\Gamma}_{z}^{(s)}W_{\varrho}^{(s)},
W_{H_{NL}}^{(s)}\}\,,  
\label{flow2}
\end{equation}
where
\begin{eqnarray}
\hat{\Gamma}_{z}^{(\pm )} & = & \frac{(1\pm \varepsilon )}
{\varepsilon }\,\pm i 
\frac{(\mathbf{n}\times \hat{\mathbf{L}})_{z}\,}{n_{z}}\, , 
\nonumber\\
& & \\
\hat{\Gamma}_{z}^{(0)} & = & \frac{1}{2\varepsilon} 
\Phi (\mathcal{L}^{2}) - \frac{1}{2}\varepsilon 
\left [ 1+2i\frac{(\mathbf{n}\times \hat{\mathbf{L}})_{z}}{n_{z}} \right ] \,  \Phi ^{-1}(\mathcal{L}^{2}) \,, \nonumber 
\end{eqnarray}
and here $i(n\times \hat{\mathbf{L}})_{z}/n_{z} =\tan \theta \
\partial _{\theta }$  (in general, an arbitrary direction can be
chosen instead of the $z$ component). The equation of motion in the
form (\ref{flow2}) appears as the so-called second-kind continuity equation~\cite{Liu:2011aa}.

\section{Dissipative quasiprobability currents on the sphere}

\subsection{Dissipative currents}

Models of dissipation address the interaction of a system with an environment, whose characteristics are encoded in its spectral density~\cite{Breuer:2003aa}. Here, we assume that the spin system is coupled to a thermal bath at temperature $T$. The resulting effective dynamics is appropriately described by the Lindblad equation~\cite{Kossakowski:1972aa,Lindblad:1976aa,Gorini:1976aa}
\begin{equation}
  \partial_{t}{\hat{\varrho}}=-i   [ \hat{H},\hat{\varrho} ] +
  \case{1}{2} \gamma \, ( \overline{n}+1)
  \hat{\Lambda}_{1}(\hat{\varrho}) +
  \case{1}{2}
  \gamma \, \overline{n} \hat{\Lambda}_{2}(\hat{\varrho}) \, ,
  \label{LE}
\end{equation}
where $\hat{\Lambda}_{1,2}$ are the superoperators 
\begin{eqnarray}
   \hat{\Lambda}_{1}(\hat{\varrho}) & = & 
  2\hat{S}_{-}\hat{\varrho}\hat{S}_{+}-
  \hat{S}_{+}\hat{S}_{-}\hat{\varrho} -
  \hat{\varrho}\hat{S}_{+}\hat{S}_{-} \, ,\nonumber \\
  & & \\
  \hat{\Lambda}_{2}(\hat{\varrho}) & = &
  2\hat{S}_{+}\hat{\varrho}\hat{S}_{-} -
  \hat{S}_{-}\hat{S}_{+}\hat{\varrho} -
  \hat{\varrho} \hat{S}_{-}\hat{S}_{+} \, ,\nonumber 
\end{eqnarray}
and $\overline{n}=[\exp (\hbar \omega_{0}/kT)-1]^{-1}$ is the average number of excitations in the bath.

In the phase-space picture, the action of the superoperators $\hat{\Lambda}_{1,2}(\hat{\varrho})$ is represented by the following differential operators
\begin{widetext}
\begin{eqnarray}
 s& = & \pm 1 \quad \left \{ 
\begin{array}{lcl}
  \hat{\Lambda}_{1} (\hat{\varrho}) & \mapsto &
  [ -\hat{\mathcal{L}}^{2} (1 \pm \cos \theta ) +
  \hat{L}_{z}^{2} - \frac{1}{\varepsilon} (2\cos \theta +
  \sin \theta \partial_{\theta}) ] W_{\varrho}^{(\pm 1)}(\Omega ), \\ 
\\ 
  \hat{\Lambda}_{2}(\hat{\varrho}) & \mapsto &
  [ - \hat{\mathcal{L}}^{2} (1 \mp \cos \theta ) +
  \hat{L}_{z}^{2} + \frac{1}{\varepsilon} (2\cos \theta +\sin
\theta \partial_{\theta}) ] W_{\varrho}^{(\pm 1)}(\Omega )\, ,
\end{array}
\right .  \nonumber \\
\\
 s& = & 0 \quad \; \; \; \left \{ 
\begin{array}{lcl}
  \hat{\Lambda}_{1}(\hat{\varrho}) & \mapsto &
  [ -\hat{\mathcal{L}}^{2}+ \hat{L}_{z}^{2} -
  \frac{1}{\varepsilon} ( \cos \theta +
  \frac{1}{2}\sin \theta \partial_{\theta} ) \Phi (\mathcal{L}^{2})  +\varepsilon ( \hat{\mathcal{L}}^{2}- \cos \theta -
  \frac{1}{2}\sin \theta \partial_{\theta} )
  \Phi^{-1} (\hat{\mathcal{L}}^{2}) ] W_{\varrho}^{(0)}(\Omega ), \\ 
\\ 
  \hat{\Lambda}_{2}(\hat{\varrho}) & \mapsto & [ - \hat{\mathcal{L}}^{2} +
  \hat{L}_{z}^{2} + \frac{1}{\varepsilon} ( \cos \theta +
  \frac{1}{2}\sin \theta \partial_{\theta} ) \Phi (\hat{\mathcal{L}}^{2})  -\varepsilon ( \hat{\mathcal{L}}^{2} - \cos \theta -\frac{1}{2}\sin \theta
\partial_{\theta} ) \Phi^{-1}(\hat{\mathcal{L}}^{2})]
W_{\varrho}^{(0)}(\Omega ) \, .
\end{array}
\right .  \nonumber
\end{eqnarray}
\end{widetext}
The dissipative phase-space dynamics of these spinlike systems has been investigated by a number of authors~\cite{Perelomov:1986ly,Carmichael:1999aa,Kalmykov:2016aa}. Interestingly, the Lindblad evolution can also be recast as a continuity equation with current operators given by
\pagebreak
\begin{eqnarray}
  \label{Jtpm}
  \hat{J}_{\theta}^{(\pm 1)} & = & \case{1}{2} \gamma
 \left[ \frac{1}{\varepsilon} \sin \theta -
\partial_{\theta} ( 1 + 2 \overline{n} \pm \cos \theta ) \right] ,  \nonumber \\
  \hat{J}_{\phi}^{(\pm 1)} &=&-\case{1}{2} \gamma
\left[ (\pm \tan \theta + ( 2 \overline{n} + 1 )
 \frac{\cos \theta}{\tan \theta}\right] \partial_{\phi};
\nonumber \\
& & \\
\hat{J}_{\theta}^{(0)} & = & \case{1}{2} \gamma
 \left[ \frac{1}{2\varepsilon} \sin \theta \,
  \Phi (\hat{\mathcal{L}}^{2} ) - ( 2 \overline{n} + 1)
 \partial_{\theta}\right . \nonumber \\
&  + & \left . \varepsilon \left( \partial_{\theta} +
 \case{1}{2}\sin \theta \right) \ \Phi^{-1}(\hat{\mathcal{L}}^{2})\right] ,  \nonumber \\
  \hat{J}_{\phi}^{(0)} &=&-\case{1}{2} \gamma
 \left [ (2\overline{n} +1) \frac{\cos \theta}{\tan \theta} -
\frac{\varepsilon}{\sin \theta}\ \Phi^{-1}(\hat{\mathcal{L}}^{2} )
 \right]  \partial_{\phi}.  \nonumber
\end{eqnarray}
In the high-temperature regime ($\overline{n}\gg 1$), all the maps ($s=\pm 1,0$) lead to the same equation of motion; viz,
\begin{equation}
\hat{\Lambda}(\hat{\varrho}) \simeq \case{1}{2} \gamma \, 
\overline{n} \left( \hat{\Lambda}_{1}+ \hat{\Lambda}_{2}\right) 
(\hat{\varrho}) \mapsto - \case{1}{2} \gamma \,\overline{n} 
\left( \hat{\mathcal{L}}^{2} + \partial_{\phi \phi} \right ) 
W_{\varrho}^{(s)}(\Omega ),
\end{equation}
and the currents take the simple form 
\begin{equation}
\label{Jdn}
\hat{J}_{\theta}^{(s)} \simeq -\gamma \,
\overline{n}\, \partial_{\theta}, \qquad 
\hat{J}_{\phi}^{(s)} \simeq -\gamma \, \overline{n} \, \frac{\cos \theta}{\tan \theta} \, \partial_{\phi} \,.
\end{equation}

In the presence of dissipation, the stagnation points may become  sinks and sources. For instance, (\ref{Jdn}) generate a single source of a free-evolving vector field at the point $( \theta =\pi /2,\phi =0 ) $ for an initial spin coherent state  centered on the $X$ axis $|\Omega = (\pi /2, 0) \rangle$. In the opposite limit of  $\overline{n} =0$ in addition to that source,
several sinks appear.

\subsection{Classical limit}

In the large spin limit, $\varepsilon \rightarrow 0$, only the Wigner current tends to the classical form. Indeed, taking into account that $\Phi (\hat{\mathcal{L}}^{2}) \simeq 2+ \mathcal{O}(\varepsilon^{2})$, the operators $\mathbf{\hat{J}}^{(0)}$  reduce  to $c$-numbers, that is, $ \mathbf{\hat{J}}^{(0)} \simeq \mathbf{v} + \mathcal{O} (\varepsilon)$, with
\begin{equation}
\mathbf{v} = 2 \varepsilon \left (
\begin{array}{c}
\displaystyle \frac{1}{\sin \theta} \partial_{\phi} W_{H}^{(0)}(\Omega ) \\
- \partial_{\theta}W_{H}^{(0)}(\Omega )
\end{array} 
\right) = 2 \varepsilon \nabla W_{H}^{(0)} (\Omega ) \, .
\end{equation}
Here, $W_{H}^{(0)} (\Omega)$ is the corresponding symbol of the Hamiltonian,
which, using (\ref{eq:symbols}), is 
\begin{equation}
  W_{H}^{(0)} (\Omega ) = \frac{1}{2\varepsilon}
  \sum_{i}a_{i}\, n_{i}+\frac{1}{4\varepsilon^{2}}
  \sum_{jk}b_{jk}\, n_{k}n_{j} \, .
\label{WH}
\end{equation}
According to (\ref{Js}), in this limit the points of the distribution $W_{\varrho}^{(0)}(\Omega )$ just follow the flow generated by $\mathbf{J}_{ \mathrm{cl}}$. In contradistinction, the $Q$ and $P$ currents have nonvanishing corrections in this semiclassical limit.

In the same limit, the dominant term in the dissipative currents (\ref{Jtpm}) is $\hat{J}_{\theta}^{(s)}$ and this yields Fokker-Planck equations. This also indicates that the main direction of the dissipative motion is towards the South pole of the Bloch sphere, representing the ground state of the system.

\section{Examples: Kerr and Lipkin-Meshkov-Glick models}

\label{sec:ex}

\begin{figure}[t]
\centering
\includegraphics[width=\columnwidth]{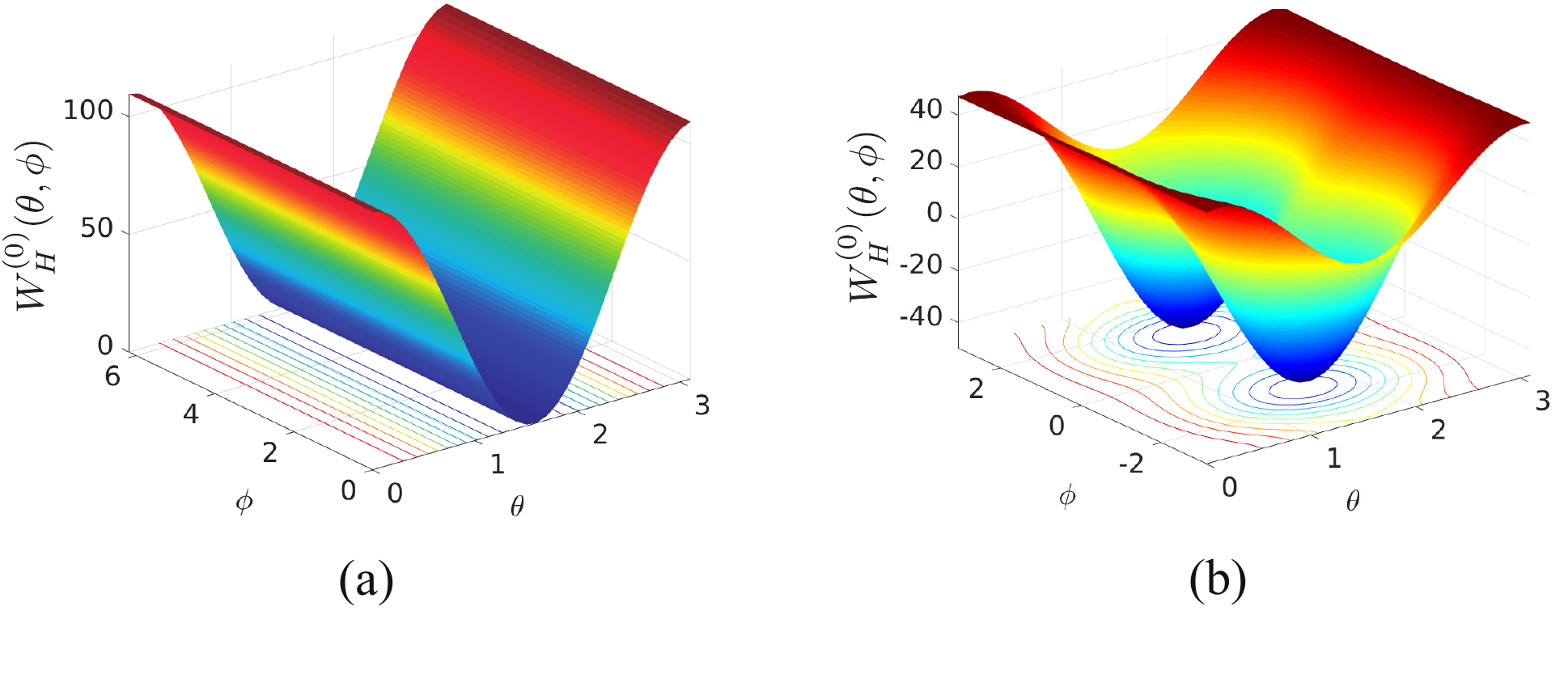}
\caption{The Wigner symbol $W^{(0)}_{H}(\Omega )$ for (a) the Kerr and (b) the LMG Hamiltonians.}
\label{fig:potenciales}
\end{figure}

\subsection{Kerr model}

The simplest quadratic Hamiltonian corresponds to the so-called Kerr medium, which is described by 
\begin{equation}
\hat{H}_{\mathrm{Kerr}}= \chi \, \hat{S}_{z}^{2} \, .  
\label{kerr}
\end{equation}
This leads to a remarkable non-Gaussian operation that has set off a lot of interest due to possible applications in a variety of fields, such as quantum nondemolition measurements~\cite{Braginskii:1968aa,Unruh:1979aa,Milburn:1983aa,Imoto:1985aa,Sanders:1989aa}, generation of quantum superpositions~\cite{Milburn:1986aa,Yurke:1986aa,Tombesi:1987aa,Gantsog:1991aa,Tara:1993aa,Luis:1995aa,Chumakov:1999aa,Rigas:2013aa},  quantumteleportation~\cite{Vitali:2000aa,Zhu:2011aa},  and quantum logic~\cite{Turchette:1995aa,Semiao:2005aa,Matsuda:2007aa,You:2012aa}.

\begin{figure*}[t]
  \centering {\includegraphics[width=1.75\columnwidth]{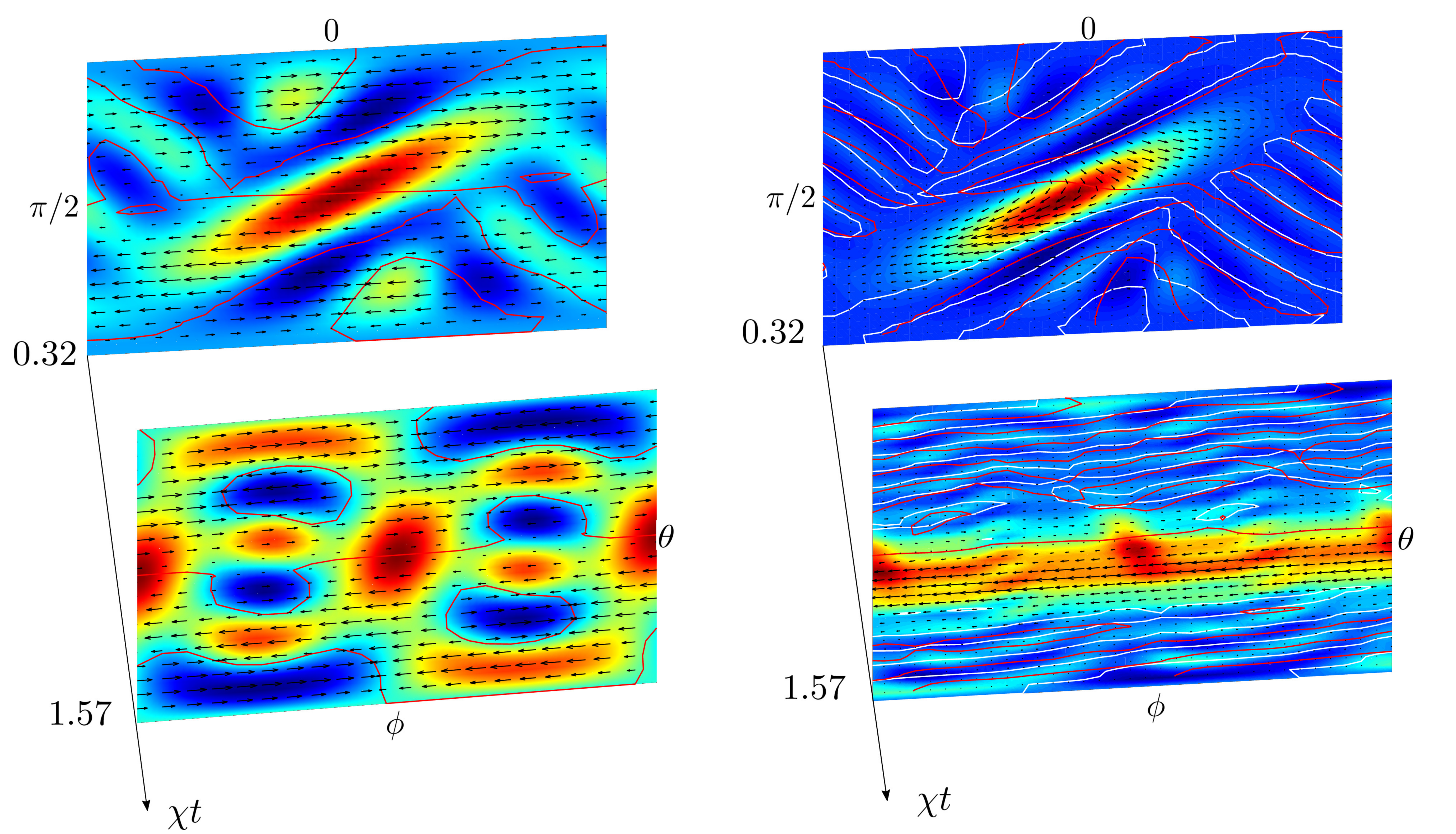}}
  \caption{Snapshots of the Wigner function corresponding to the Kerr Hamiltonian (\ref{kerr}) and the initial spin coherent state $|\Omega = (\pi /2,0) \rangle $ at the best squeezing time $\chi t=0.32$ and two-component Schr\"{o}dinger cat time $ \chi t=\pi /2$. Left panel, unitary evolution; right panel, dissipative evolution with $\gamma =0.015$. Red and white lines are stagnation lines for $J_{\theta }$ and $J_{\phi }$, respectively.  The pseudocolor encodes the Wigner function and the black arrows indicate the direction and strength of the flow. We have used $S=10$.}
  \label{fig:kerr}
\end{figure*}

As shown in Fig.~\ref{fig:potenciales}, the Wigner symbol of $\hat{H}_{\mathrm{Kerr}}$ has the aspect of a valley centered at $\theta =\pi /2$. The unitary currents have only one nonzero component, ($\hat{J}_{\theta }^{(s)}=0$), and they are
\begin{eqnarray}
\hat{J}_{\phi }^{(0)} &=&\sin \theta \left[ \tfrac{1}{2 \varepsilon}\cos \theta \;\Phi (\hat{\mathcal{L}}^{2}) - \tfrac{1}{2}\varepsilon (\cos \theta
+ 2\sin \theta \partial _{\theta })\Phi ^{-1}(\hat{\mathcal{L}}^{2})\right]
\,,  \nonumber  \label{Jt0} \\
\hat{J}_{\phi }^{(1)} &=&\sin \theta \left[ 2(S+1)\cos \theta +\sin \theta \partial _{\theta }\right] \,, \\
\hat{J}_{\phi }^{(-1)} &=&\sin \theta \left[ 2S\cos \theta -\sin \theta \partial _{\theta }\right] .  \nonumber
\end{eqnarray}

A peculiarity of these currents is the existence of stagnation lines, where $J_{\phi}^{(s)}(\Omega )=0$ (see also the discussion in \cite{Yang:2019aa}). When dissipation is included, both components are present. In Fig.~\ref{fig:kerr} we plot two snapshots of the Wigner function for an initial coherent state (\ref{CS}) located at the equator $\Omega = (\pi /2,0) $ both for unitary and dissipative dynamics. The first one is at the best squeezing time ($\chi t\sim S^{-2/3}$),  where the maximum value of the spin squeezing, evaluated as the normalized minimum fluctuation of spin components $\Delta^{2} \mathbf{s}(t)$ on the tangent plane, orthogonal to the initial mean spin vector $\mathbf{n} =\langle \hat{\mathbf{S}}(t)\rangle$ ($\mathbf{s\cdot n}=0$) is achieved. The second one is  at the two-component cat time ($\chi t=\pi /2$)~\cite{Agarwal:1997aa}, when the state becomes a superposition of two spin coherent states ($|\Omega =(\pi /2,0)\rangle + |\Omega =(\pi /2,\pi )\rangle $).

The currents (\ref{Jt0}) apparently generate a fast motion of initial distributions to the south pole of the Bloch sphere; i. e., a decay into the ground state $|S,-S\rangle $. Nonetheless, distributions localized inside the potential valley move quite slowly into the south pole, so that even some quantum interference effects like residual Schr\"{o}dinger cat states can be observed. In Fig.~\ref{fig:kerr} we see that the interference pattern is partially destroyed for times $\chi t\sim 1$, as expected, but the distribution is still mainly concentrated inside the valley, in spite of the strong dissipation.

This can be understood by taking into account that for $S=10$ at the minimum of the valley $W_{H}^{(0)}(\theta_{\min },\phi )\simeq 0.125$ and the energy fluctuation in the state $|\Omega = (\pi /2,0) \rangle $ is $\Delta H\simeq 6.8$, while at the south pole $W_{H}(\theta =\pi ,\phi )\simeq 109.75$. Thus, the distribution should overcome a significant potential barrier in order to reach the south pole, which slows down the decay of equatorially localized distributions to the ground state.

\begin{figure}[b]
  \centering {\includegraphics[width=.90\columnwidth]{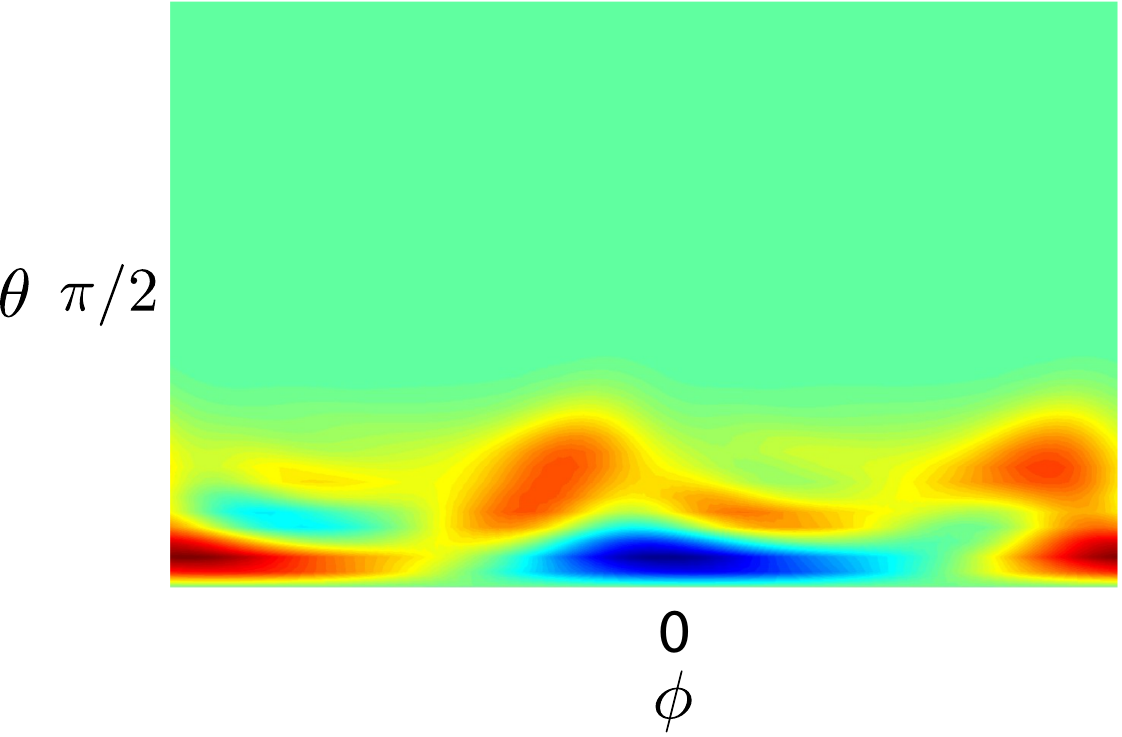}}
  \caption{Pseudocolor plot of the Wigner function for the initial state $|\Omega = (3\pi /4,0)\rangle $ at the two-component cat time in case of dissipative evolution ($\gamma =0.015$) with $S=10$.} 
  \label{fig:kerrc}
\end{figure}

It is interesting to stress that, in contrast to the previous behavior, when the distributions are localized below the equator they rapidly slide toward the south pole.  This can be readably observed in Fig.~\ref{fig:kerrc}, where the Wigner function of an initial spin coherent state with $\Omega = (3\pi /4,0)$ at the two-component cat time, $\chi t=\pi /2$, is shown.

\subsection{Lipkin-Meshkov-Glick model}

The Lipkin-Meshkov-Glick (LMG) model~\cite{Lipkin:1965aa,Meshkov:1965aa,Glick:1965aa} was originally proposed to deal with phase transitions in the nuclei. The model captures well the physics of two-mode Bose-Einstein condensates~\cite{Ulyanov:1992aa} and Josephon junctions~\cite{zibold:2010aa,Julia:2012aa}. In the language of spin operators the LMG Hamiltonian can be written as
\begin{equation}
  \hat{H}_{\mathrm{LMG}} = - h \, \hat{S}_{x}+\frac{\lambda}{2(2S+1)}
  ( \hat{S}_{z}^{2}-S_{y}^{2} ) \, .
  \label{lipkin}
\end{equation}
For different values of the parameter $\lambda $,  the associated classical symbol has either one or two minima. Here, we take $\lambda \sim S$, which corresponds to a double-well potential, as shown in Fig.~\ref{fig:potenciales}. The minima are located at $(\theta_{\min}=\pi /2, \phi_{\min}=1.459)$ and $(\theta_{\min}=\pi /2, \phi_{\min}=-1.459 )$, separated by a local maximum with a saddle point at $(\theta_{s}=\pi /2,\phi_{s}=0)$. The LMG current operators are significantly more involved than (\ref{Jt0}), and have the form in accordance with  the general expressions~(\ref{Jops}).

We consider two dynamical regimes:

\subsubsection{Stable motion}
\begin{figure*}[t]
  \centering {\includegraphics[width= 1.45\columnwidth]{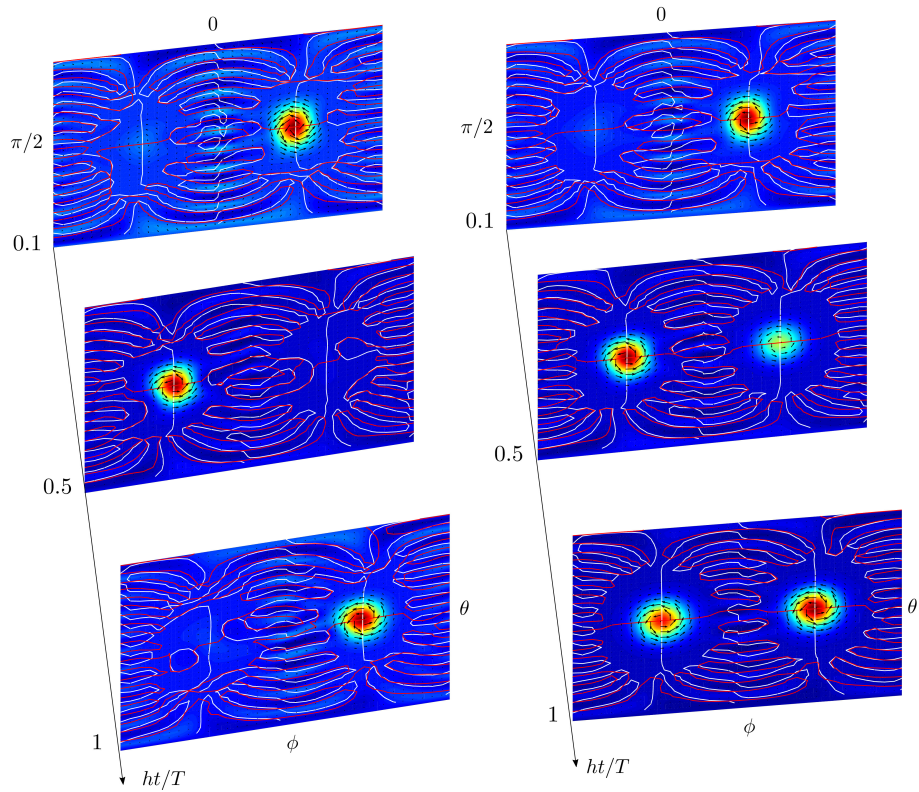}}
  \caption{Snapshots of the Wigner function corresponding to the LMG Hamiltonian (\ref{lipkin}) for the stable case. The initial spin coherent state is $|\Omega = (\pi /2,-1.459) \rangle $ at the times $h t/T=0.1$ (back), $h t/T= 0.5$ (middle) and $h t/T=1$ (front), where $T$ here is the period of oscillation between the two wells. Left panel,  unitary evolution; right panel,  dissipative evolution with $\gamma =1\times 10^{-7}$.  Red and white lines are stagnation lines for $J_{\theta}$ and $J_{\phi}$, respectively. In both cases the total spin is $S=10$.}
  \label{fig:stable}
\end{figure*}
\begin{figure}[b]
\centering
\includegraphics[width=0.80 \columnwidth]{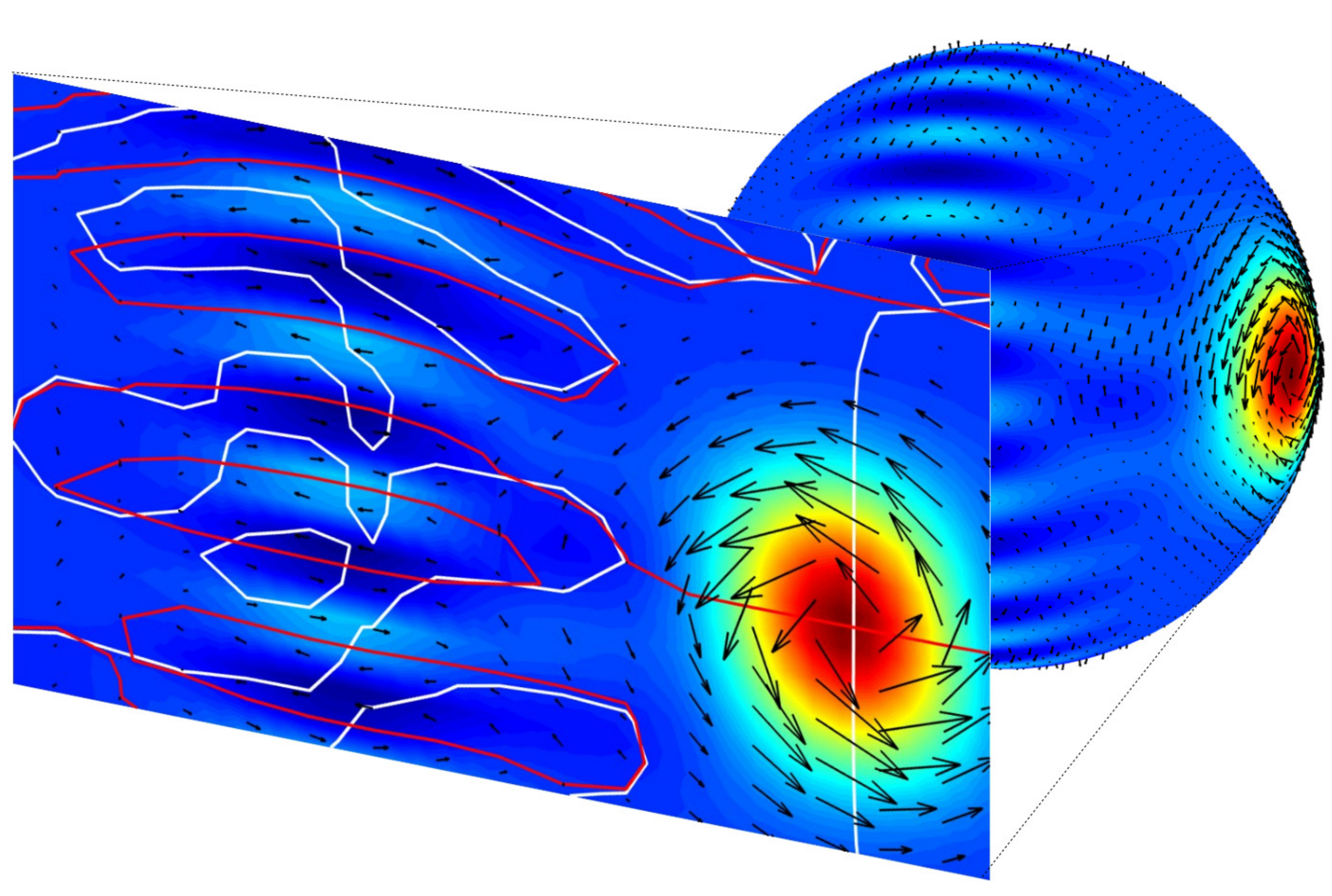}  
\caption{Zoom of the Wigner function and the distribution of currents at $h t/T=0.1$ in the vicinity of the saddle point for the LMG Hamiltonian for the stable unitary evolution. The initial stat is the same as in Fig.~\ref{fig:stable}.}
\label{fig:esfera}
\end{figure}

We take an initial spin coherent state with $\Omega= (\pi /2,1.459 )$ located inside one of the potential wells (actually, the right one). The classical stable motion corresponds to oscillations inside that well. Yet, due to tunnelling, the state is slowly transfered to the other well. In Fig.~\ref{fig:stable} we plot snapshots of the Wigner current at some representative times. The dynamics of the tunnelling as well as the formation of the corresponding interference patterns can be clearly appreciated.

The current directions and its intensity distribution provide nontrivial information about tunnelling dynamics indicating the main paths of the state transfer. In particular, phase-space currents explicitly show the spatial distribution of the quasiprobability flow in classically forbidden areas, where the standard density current $\mathbf{j}\propto \I (\psi ^{\ast}\mathbf{\nabla }\psi )$ is zero.  The red/white lines correspond to zero lines of $J_{\phi }^{(0)}/J_{\theta }^{(0)}$, respectively, and their intersections reveals the position of the stagnation points.

In Fig.~\ref{fig:esfera} we plot a magnification of the vicinity of the saddle point, $\phi_{s}=0$ [at the same time as the first plot in Fig.~\ref{fig:stable}]. The tunnelling flow in both directions is manifest.

\begin{figure}[b]
\centering
\par
\includegraphics[width=0.80 \columnwidth]{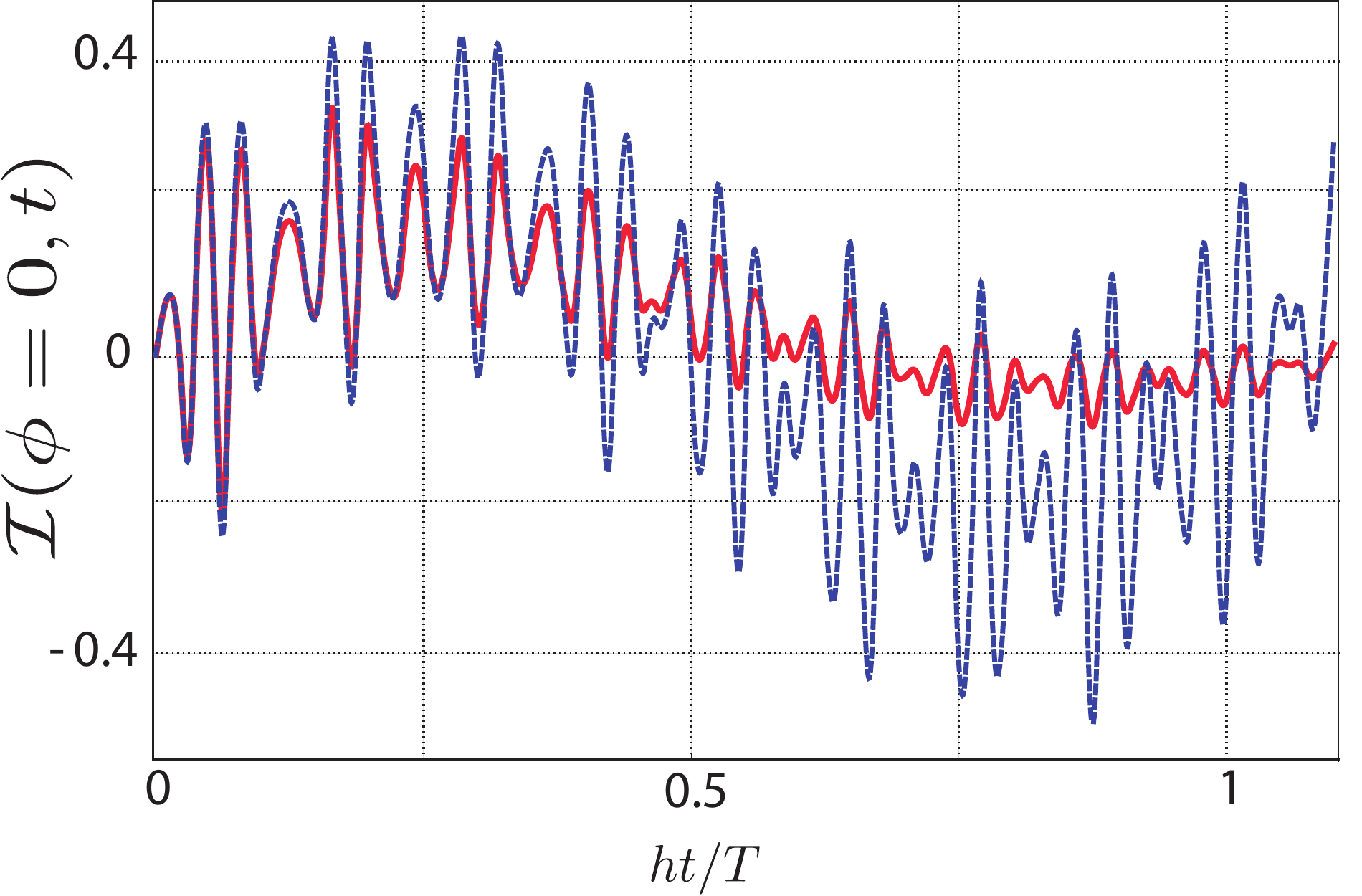}
\caption{The flow $\mathcal{I} ( \phi =0, t)$ (\ref{I}) for the same unitary (blue line) and dissipative (red line) dynamics as in Fig.~\ref{fig:stable} in terms of the dimensionless time $ht/T$.} 
\label{fig:intphi}
\end{figure}

\begin{figure*}[t]
  \centering {\includegraphics[width=1.65\columnwidth]{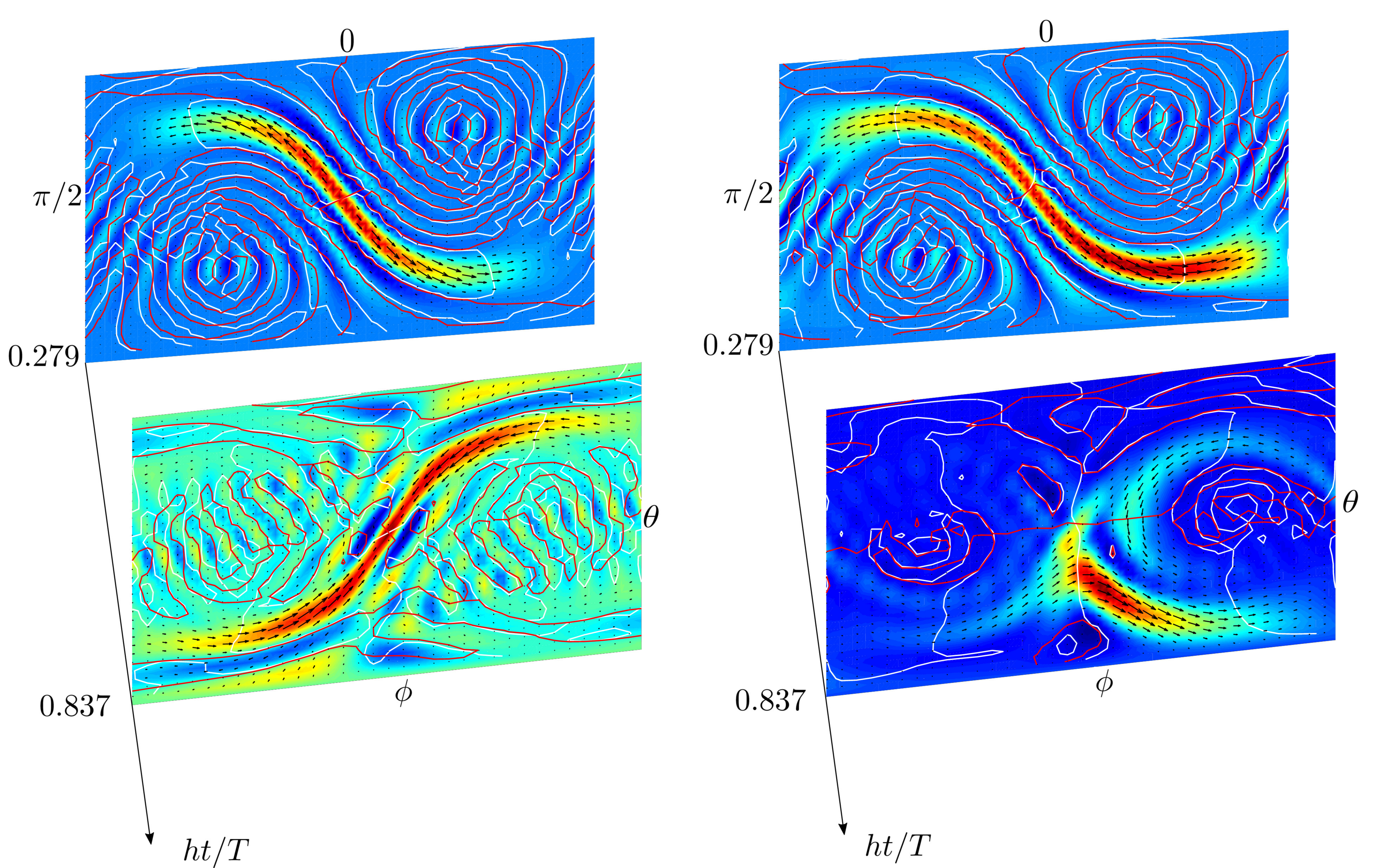}}

  \caption{Snapshots of the Wigner function corresponding to the LMG Hamiltonian for the unstable case. The initial spin coherent state is $|\Omega = (\pi /2,0) \rangle $ and the dimensionless times $h t=0.279$ (back) and $h t= 0.837$ (front). Left panel, unitary evolution; right panel, dissipative evolution with $\gamma =0.05$. Red and white lines are stagnation lines for $J_{\theta}$ and $J_{\phi}$, respectively. Again, the total spin is $S=10.$}
  \label{fig:unstable}
\end{figure*}

The counterclockwise vector field in the vicinity of the stable point $ (\theta_{\min}=\pi /2,\phi_{\min}=-1.459)$ extends into the classically forbidden region, generating a left-to-right flow with the highest intensity in the region $\theta <\pi /2$ (above the equator). The tunnelling flow freely crosses the stagnation (white) line $J_{\theta}^{(0)}(\Omega )=0$. The flow in the opposite direction is mainly below the  equator, as one can see in Fig.~\ref{fig:esfera}.

When a moderate dissipation is present, the tunnelling becomes slower, in agreement with general considerations~\cite{Caldeira:1981aa}. The interference pattern is largely destroyed. Nonetheless, in spite of the very long transfer times from one well into another, the distribution does not show any fingerprint of decay to the ground state at the \emph{half-period} of motion (when the distribution has passed to another minimum). The reason of such a behavior is the same as in the Kerr medium: to reach the south pole following the lines of the dissipative current, a distribution initially localized in the minimum of the potential should overcome a potential barrier, which is significantly higher than the local maximum: $ W_{H}^{(0)}(\theta_{\min},\phi_{\min})=-47.567$; the energy fluctuation in the state $|\Omega = (\pi /2,1.459) \rangle $ is $\Delta H \sim 2.996$, $W_{H}(\theta _{s},\phi_{s})=-10.488$, while $W_{H}(\theta =\pi ,\phi )=46.982$, for our case of $S=10 $.

The dissipative evolution for larger times drastically differs from the unitary one: while the Hamiltonian evolution is quasiperiodic and the distribution \emph{oscillates} between the potential wells, the dissipation does not allow to the transfer of the whole distribution to the other well, and the \emph{inverse tunnelling} back to the original well is significantly suppressed in comparison with the unitary case. Actually, the decoherence in a two-well tunnelling acts as a viscous medium, in the sense that it leads to a phase-space equilibration at long-times, when the initial quasiprobability becomes equally distributed between the wells in the form of an incoherent superposition.

Useful information about the tunnelling is provided by the integral flow at the line $\phi =0$ (which separates the potential wells), 
\begin{equation}
\mathcal{I}(\phi =0,t) = \int d\theta \sin \theta \; 
 J_{\phi}^{(0)}(\theta ,\phi =0, t) \, ,
\label{I}
\end{equation}
which typifies the dominant direction of the propagation at a given time.

In Fig.~\ref{fig:intphi} we plot the flow (\ref{I}) for one \emph{period} of the tunnelling oscillation. During this time the initial distribution is transferred to the other potential minimum and returns back. One can observe that the direction of the Hamiltonian evolution changes from right-to-left to left-to-right when the distribution is completely transferred from the right to the left potential well. The flow in the presence of dissipation is significantly smaller in the second half-period of motion.

\subsubsection{Unstable motion.}

Next, we consider the initial spin coherent state $|\Omega = (\pi /2,0) \rangle $ centered at the saddle point (the classical separatrix). The directions of the current clearly indicate the hyperbolic nature of the stagnation point $(\theta_{s}=\pi /2,\phi_{s}=0)$. Quantum instability is reflected in a separation of the initial distribution into two symmetric pieces moving toward the minima, according to the current direction, with a subsequent formation of a complex interference picture. This can be seen in Fig.~\ref{fig:unstable}, where the snapshots of the Wigner function along with the corresponding current lines are plotted. As the state is initially at the local maxima, the decay to the ground state is quite fast in the presence of dissipation. The distribution clearly tends to the south pole along the current lines at times approximately corresponding to the half-period of oscillations in the Hamiltonian case, in sharp contrast with the stable situation.

\section{Conclusions}

In summary, we expect to have provided compelling evidence demonstrating that the quantum currents $\mathbf{J}^{(s)}(\Omega |t)$are a useful tool for the analysis of the evolution in phase space. Indeed, the spatial distribution of the quantum current allows one to visualize the main directions of propagation of the distribution. Our analytic current operators explicitly underlines the strong differences between the Wigner and the $Q$ and $P$-currents for SU(2) quadratic Hamiltonians: while for $Q$ and $P$ the quantum effects are generated by the gradient operator, the Wigner current also involves action of the Laplace operator, which leads to a
significantly more involved phase-space interference patterns.

The effect of dissipation is twofold: it destroys the interference and generates a flow towards the south pole. Nevertheless, as we have seen in the example of the stable LMG evolution, the impact of the decoherence on a given Hamiltonian dynamics depends essentially on the location of the initial distribution.

It is worth noting that in multi-spin case, when the classical phase space is a direct product of several two-spheres, the geometrical representation of the currents (\ref{Js}) would not be directly possible, but the analytical properties of the stagnation points (e.g. the winding numbers) still can provide a useful information about the character of nonlinear evolution.

\section*{Acknowledgments}

This work is partially supported by the Grant 254127 of CONACyT (Mexico). L.L.S.S acknowledges the support of the Spanish MINECO (Grants FIS2015-67963-P and PGC2018-099183-B-I00).

%

\end{document}